\documentclass[sigconf]{acmart}

\AtBeginDocument{%
  }

\setcopyright{acmlicensed}
\copyrightyear{2024}
\acmYear{2024}
\acmDOI{XXXXXXX.XXXXXXX}

\usepackage{tabularx}

\usepackage{graphicx}      
\usepackage{multirow}
\usepackage{caption}
\usepackage{subcaption}
\usepackage{soul}
\usepackage{url}

\graphicspath{{./figs/}}

\begin{document}

\title{Design and Evaluation of Camera-Centric\\Mobile Crowdsourcing Applications}


\author{Abby Stylianou}
\affiliation{%
  \institution{Saint Louis University}
  \city{Saint Louis}
 \state{MO}
 \country{USA}}
\email{abby.stylianou@slu.edu}

\author{Michelle Brachman}
\affiliation{%
  \institution{IBM Research}
  \city{Cambridge}
 \state{MA}
 \country{USA}}
\email{michelle.brachman@ibm.com}

\author{Albatool Wazzan}
\affiliation{%
  \institution{Temple University}
  \city{Philadelphia}
 \state{PA}
 \country{USA}}
\email{albatool.wazzan@temple.edu}

\author{Samuel Black}
\affiliation{%
  \institution{Temple University}
  \city{Philadelphia}
 \state{PA}
 \country{USA}}
\email{sam.black@temple.edu}

\author{Richard Souvenir}
\affiliation{%
  \institution{Temple University}
  \city{Philadelphia}
 \state{PA}
 \country{USA}}
\email{souvenir@temple.edu}

\renewcommand{\shortauthors}{Stylianou et al.}

\begin{abstract}
The data that underlies automated methods in computer vision and machine learning, such as image retrieval and fine-grained recognition, 
often comes from crowdsourcing.
In contexts that rely on the intrinsic motivation
of users, we seek to understand how the application design affects a user's 
willingness to contribute and the quantity and quality of the data they capture. 
In this project, we designed three versions of a camera-based mobile crowdsourcing 
application, which varied in the amount of labeling effort requested of the user 
and conducted a user study to evaluate the trade-off between the level of 
user-contributed information requested and the quantity and 
quality of labeled images collected. The results suggest that higher 
levels of user labeling do not lead to reduced contribution. 
Users collected and annotated the most images 
using the application version with the highest requested level of 
labeling with no decrease in user satisfaction. In preliminary experiments, the additional
labeled data supported increased performance on an image retrieval task.
\end{abstract}


\maketitle

\section{Introduction}
Modern machine learning applications rely on example data, which is often acquired and labeled by people. 
The steady increase in the 
performance of these methods has been fueled by a corresponding growth in the availability of 
high-quality labeled data.
For the case of images, 
mobile devices are the primary modality for capturing and submitting relevant photographs. 
In addition to the images themselves, these camera-centric mobile applications often request user-provided labels or annotations. 
Designing these applications effectively can be quite challenging, especially in the context of citizen science~\citep{citizen_science} applications, 
which rely on the intrinsic motivation of users to contribute to projects via crowdsourcing. The design of mobile crowdsourcing applications 
should ensure that the image capture and label process does not discourage users, yet still results in effectively labeled data. 

\begin{figure}
    \centering
     \includegraphics[width=.32\linewidth]{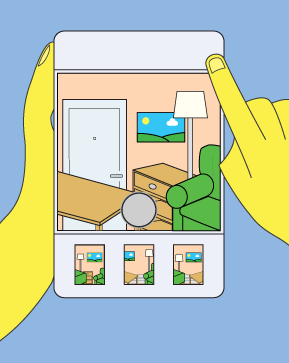}
      \includegraphics[width=.32\linewidth]{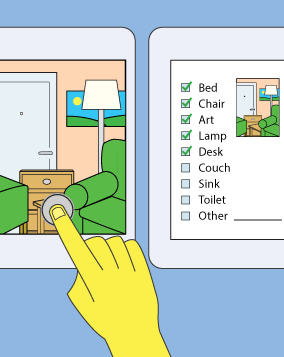}
      \includegraphics[width=.32\linewidth]{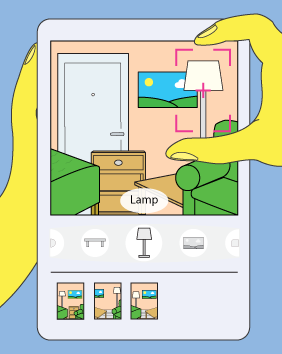}
        \caption{Image retrieval systems are dependent on high quality, well labeled data. In this study, we designed and evaluated three variants of a camera-centric mobile application for crowdsourcing data for such a system. The variants require varying amounts of image labeling from the user: (from L to R) none, weak (naming objects), and strong (naming and locating objects).}
        \label{fig:teaser}
\end{figure}

There is a general consensus that the on-boarding process, or amount of effort 
required for a user to \textit{start} using a crowdsourcing application should be minimized 
(e.g., by not requiring user signups~\citep{Jay2016} or keeping tasks small and easy to 
understand~\citep{Eveleigh2014}). 
However, it is not clear
that this should be a universal guideline. In the case of crowdsourcing applications, 
minimizing user effort may limit the type and/or amount of data that they might actually 
be willing to contribute. Recent work suggests that requesting more effort from contributors does not actually 
lead to lower engagement or user satisfaction, as in the context of audio labeling~\citep{Cartwright2019}. There has been little research investigating this phenomenon for the increasingly-ubiquitous
class of camera-centric mobile applications, such as popular citizen applications used for bird watching~\citep{inaturalist,ebird} or environmental studies~\citep{goeau2013pl}.


In this work, we compare three designs of a camera-centric mobile application, as shown in Figure~\ref{fig:teaser}, which 
differ based on the type of information requested: 
\begin{description}
    \item [Unlabeled] For the baseline method, the user takes a picture, and no additional information beyond the scene captured in the image is required. 
    \item [Weakly Labeled] ``What is in the image?'' In addition to image capture, the user is asked to name (or classify) the objects contained in the scene  by selecting from a pre-defined list.
    \item [Strongly Labeled] ``What is in the image and where is it?'' The user is asked to identify the location of particular objects in the image by either changing the focus area in the application to outline a particular object or capturing the image in a way that the object of interest is within the focus area boundaries. 
\end{description}
This categorization aligns with popular paradigms in machine learning: unsupervised learning processes unlabeled training data, 
weakly supervised data involves training data whose annotations are limited in some manner, 
and strongly supervised (or more commonly, simply supervised) learning makes use of fully annotated training data. 
We conducted a user study to evaluate the trade-offs between the requested level of labeling, the quantity and quality of 
labeled images collected by participants, and user satisfaction with the different application variants. 
The results of this study suggest that, for the case of camera-centric crowdsourcing mobile applications, higher levels
of labeling effort do not lead to less engagement or user satisfaction. In fact, we observed the opposite; users collected 
and annotated the most images using the application version with 
the highest level of labeling with no decrease in user satisfaction. These findings could help to inform the design and 
implementation of of mobile crowdsourcing applications. 

\section{Background}

Crowdsourcing leverages the knowledge and understanding of the crowd to generate, annotate, and/or analyze data. 
Commonly, such tasks are outsourced to online marketplaces such as 
Amazon Mechanical Turk (AMT)\footnote{\url{https://mturk.com}}, where 
a distributed collection of workers are paid a small fee to complete 
a well-defined task~\citep{Kittur2008,Sorokin2008UtilityDA}. Other campaigns, which
fall under the umbrella of citizen science or participatory sensing, 
rely on volunteers, who often participate out of their own personal 
or scientific interest~\citep{citizen_science}. Recent work suggests 
that these intrinsic motivations (e.g., altruism, moral obligation, sense of social good, curiosity) can be as compelling as financial compensation~\citep{Law2016,Qaurooni2016,restuccia2016incentive,Rogstadius_2021}.

\begin{figure}
    \centering
    \includegraphics[width=.25\linewidth]{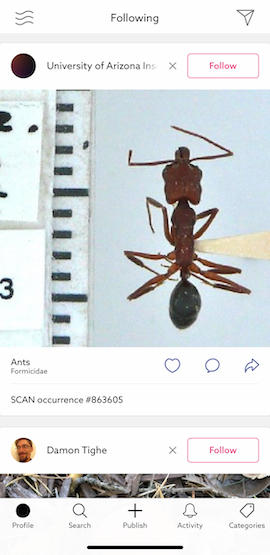}
    \hfill
    \includegraphics[width=.25\linewidth]{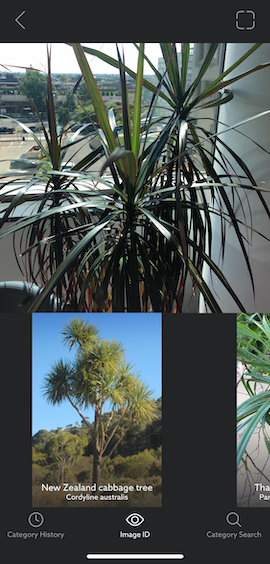}
    \hfill
    \includegraphics[width=.25\linewidth]{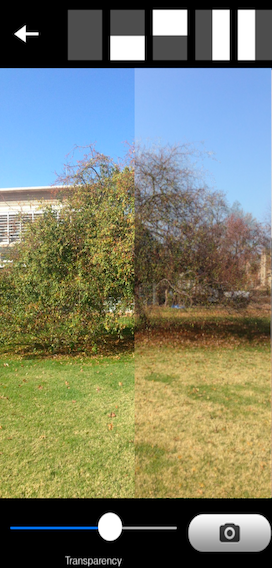}
    \caption{Fieldguide, iNaturalist, and rePhoto are examples of camera-based mobile applications that require (a) no, (b) weak, and (c) strong labeling, respectively, from the users.}
    \label{fig:cvApps}
\end{figure}

In the fields of computer vision and machine learning, there is a long history of leveraging human expertise to provide training data for automated, learning-based algorithms~\citep{crowdsourcing_cv}. This includes both the collection of imagery, as well as task-specific annotations providing information about the images (e.g., scene classification labels~\citep{places}, object classification labels~\citep{imagenet_cvpr09,Deng2014,pascal_voc,Krause2013CollectingAL,wah2011}, object bounding boxes~\citep{ILSVRC15,AAAIW125350}, per-pixel image 
labels~\citep{arbelaez2010contour,decomposing_iccv2009,mscoco,labelMe}, and image and object attribute labels~\citep{Farhadi09describingobjects,learning_visual_attibutes,unseen_classes,relative_attributes,attribute_learning}).

\newlength{\verLength}
\setlength{\verLength}{1.1in}
\begin{figure*}
    \centering
     \begin{subfigure}[b]{0.19\textwidth}
     \centering
    \includegraphics[width=\verLength]{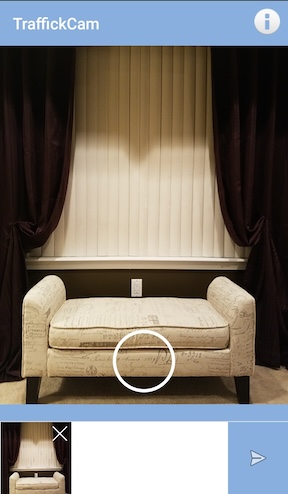}
    \caption{Unlabeled (UL)}
    \label{fig:3versions-a}
    \end{subfigure}
     \begin{subfigure}[b]{0.38\textwidth}
     \centering
      \includegraphics[width=\verLength]{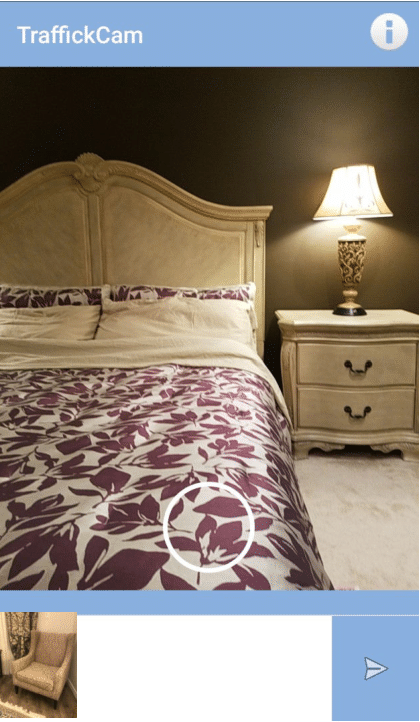}
        \includegraphics[width=\verLength]{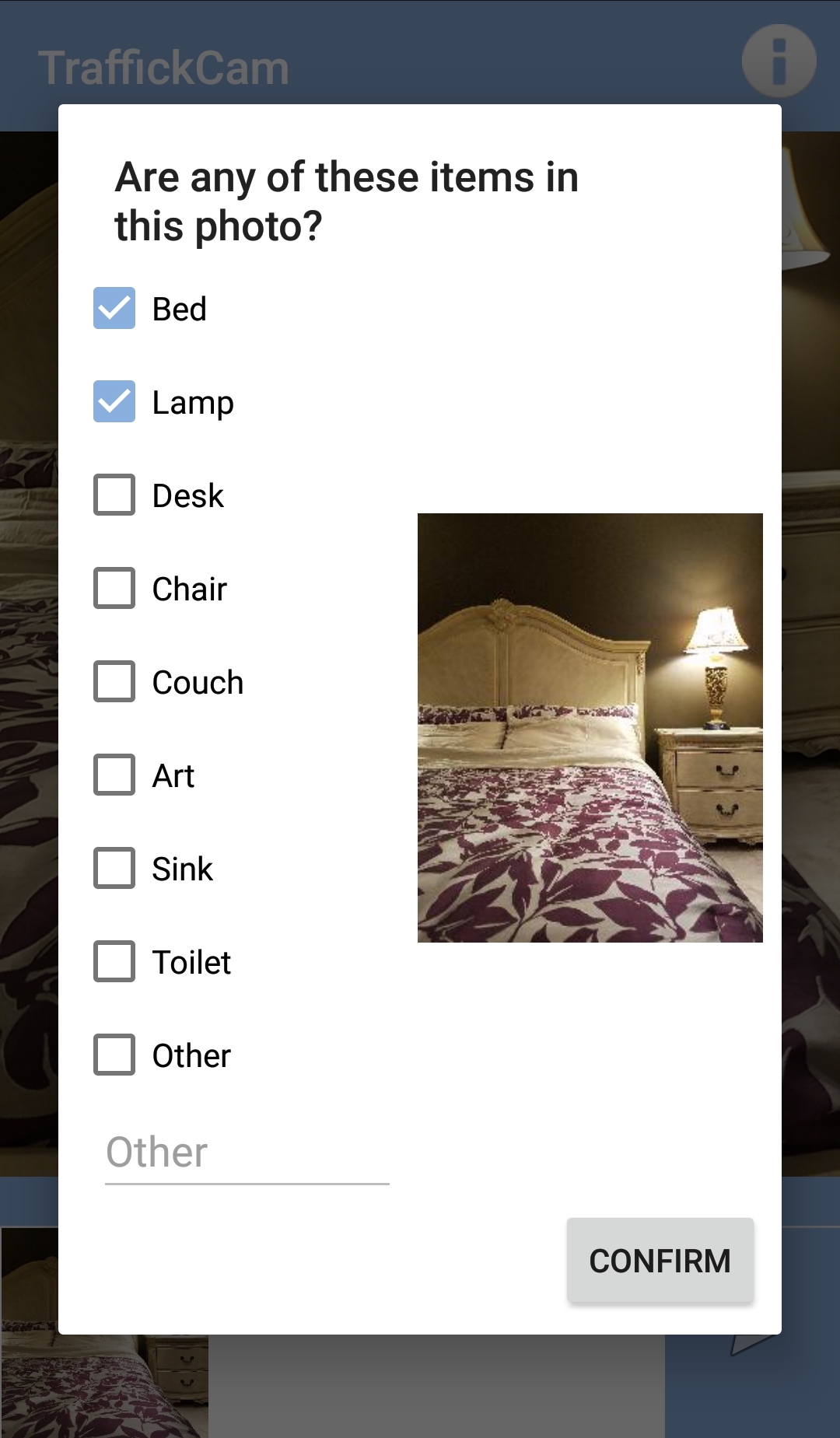}
    \caption{Weakly Labeled (WL)}
        \label{fig:3versions-b}
    \end{subfigure}
     \begin{subfigure}[b]{0.38\textwidth}
     \centering
      \includegraphics[width=\verLength]{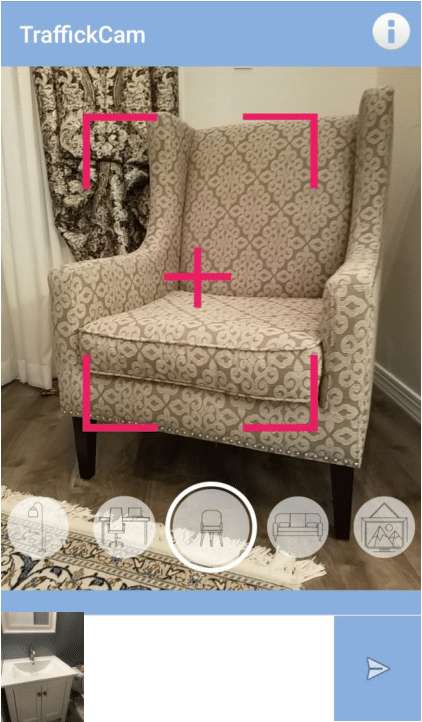}
        \includegraphics[width=\verLength]{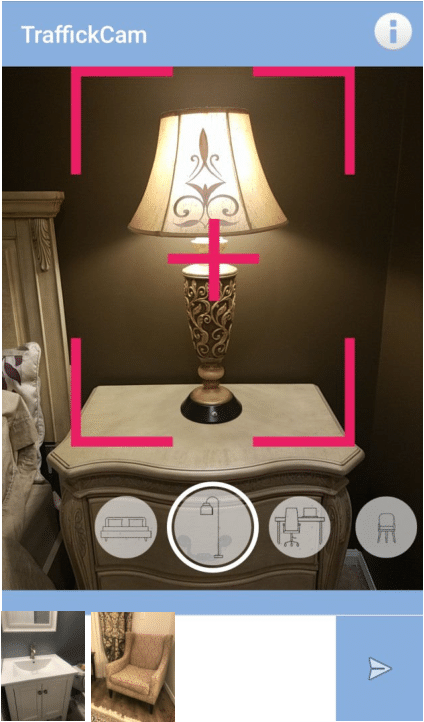}
        \caption{Strongly Labeled (SL)}
            \label{fig:3versions-c}
    \end{subfigure}
    \caption{Design variants of the camera-based mobile application for capturing 
images and identifying objects in the scene.}
    \label{fig:3versions}
\end{figure*}

There are a number of camera-centric mobile applications designed to collect images and annotations from users. WildMe's Flukebook application\footnote{\url{https://www.flukebook.org/}} allows users to submit images of whales and dolphins in order to identify particular animals and also estimate population sizes and motion patterns. In~\cite{parham2017animal}, there is a similar census of zebras and giraffes using over 50,000 user-contributed images. Fieldguide\footnote{\url{https://fieldguide.ai}} applies deep convolutional neural networks to predict the species in user contributed imagery, and relies on experts to find errors and update the predictive models (Figure~\ref{fig:cvApps} (left)). The Picture Post application\footnote{\url{https://picturepost.unh.edu}} allows users to identify locations where a 3D printed ``picture post'' has been set up to capture aligned imagery for time-series studies. In these examples, the user only contributes the captured images; the data is unlabeled. 

While the majority of available camera-centric applications fall under the unlabeled paradigm, there are some camera-centric applications that request more effort beyond image capture. Weak labeling applications such as iNaturalist~\citep{inaturalist} (Figure~\ref{fig:cvApps} (middle)), IveGotOne~\citep{wallace2014identifying} and eBird~\citep{ebird} request not only the picture, but an annotation of what plant or animal was captured. In the PlantNet application~\citep{goeau2013pl}, users first provide a plant photo, annotate the parts (e.g., leaves, flowers, stems, etc.), and then asked to validate the prediction of the plant provided by a pre-trained machine learning model. BScanner~\citep{accessibility_maps} is a mobile application to crowdsource an image dataset of outdoor locations annotated by their accessibility to aid, for example, those using a wheelchair or the visually impaired. Users provide not only images, but also identify observed accessibility problems using a dropdown menu. These applications often provide a predefined list of choices to simplify user input. In this category, there are two main operations (capturing and labeling) and, depending on the task, could be prescribed in either order: \emph{label-then-capture} or \emph{capture-then-label}. For \emph{label-then-capture} applications, each image typically contains a single object of interest. The \emph{capture-then-label} model is more amenable to labeling more than one object per image. No matter the order of operations or number of objects per image, weakly labeled images are characterized by metadata which includes the name(s) or type(s) of object(s) of interest captured in the image.

Strong labeling can take on many forms. Obtaining the classification and location of an object in an image can be accomplished by requesting the user to capture the image in a particular manner or providing annotation tools after the image has been captured. RePhoto~\citep{rephoto} (Figure~\ref{fig:cvApps} (right)) is one such application, which presents to users a semi-transparent overlay and asks them to align the camera to capture as similar of an image as possible. Other types of strong labeling tasks involve having the user provide further details about the object or scene (e.g., object details, weather, other explanations). The SeeClickFix application allows community members to report problems in their community such as potholes or illegal dumping of trash, along with photos, responses to prompts and text descriptions of the problem shown in the image in order to help improve their community. Other examples include citizen science projects hosted by Zooniverse\footnote{\url{https://www.zooniverse.org/projects}}, such as the Galaxy Zoo, which asks users to describe, identify, and differentiate between different galaxies captured by telescopes and the Wild Gabon project, which asks users to draw bounding boxes around a variety of different species in images from Gabon.

This organization (unlabeled, weakly labeled, strongly labeled) provides a categorization for a wide variety of camera-centric applications. Each category could be subdivided further based on finer-grained design decisions. To ground the evaluation of different camera-centric mobile application paradigms, we consider an application designed to collect data for indoor scene identification, specifically images from hotel rooms.

\section{Application Design}

Our study is centered on a mobile application designed to provide data to aid in human trafficking investigations~\citep{hotels50k},
where images are often important pieces of evidence, as they often contain clues about where victims have been trafficked. Much of this photographic evidence is captured in
hotel rooms. The mobile application allows travelers who want to help combat human trafficking to contribute photos of their hotel room. These images are added to a database that also includes images from publicly available travel websites (e.g., Expedia, TripAdvisor).
This database of images serves as training data for 
a learning-based reverse image search engine where investigators can submit photographic evidence in order 
to determine the hotel where a victim was photographed. As with most machine learning systems, additional training data both in terms of quantity and variability is generally beneficial for improving performance. 

The original version of the application fell into the unlabeled paradigm; users are asked to provide images of hotel rooms without any constraints or 
additional annotations. Given that hotel rooms generally contain a collection of 
common objects (e.g., bed, lamp, chair), images with object labels would increase
the utility of the AI platform to investigations by supporting more complex object-centric queries by the users of the platform. For example, an investigator may notice a particularly unique lamp in a victim image and want to search for any images with visually similar lamps (regardless of the other objects in the image).

The application could be extended to incorporate object labeling from the engaged user base already providing images to support these types of investigations. However, attracting and maintaining contributors is an important consideration for any crowdsourcing application, so new designs should not decrease motivation or interest in contributing.  

To better understand this issue, we designed three variants of the mobile application for capturing images of hotel rooms and identifying objects in the scene. We will refer to these as: Unlabeled (UL), Weakly Labeled (WL), and Strongly Labeled (SL). In this section, we describe our design decisions.


The application launches with an introduction to the application and its purpose. The next screen shown to the user provides instructions about how to use the specific version of the interface.  After viewing the introduction and instructions, the user will capture and (depending on the version) label images using the interfaces seen in Figure~\ref{fig:3versions}. After the user is satisfied with the images captured, the application requests hotel information while uploading images and metadata to a server in the background. 

\paragraph{Unlabeled (UL)} The user is 
instructed to take pictures in the hotel room without reference to 
any specific objects in the scene or 
options for additional labeling, as shown in Figure~\ref{fig:3versions-a}. The user can capture images 
and/or delete captured images. Similar to other applications in this class, the 
collected 
data includes only the 
image data and automatically collected metadata (e.g., date, time, GPS location). 

\paragraph{Weakly Labeled (WL)} The user captures a photo and labels the objects in the photo from a list. The user first captures an image in 
the same manner 
as in the UL version. Post-capture, a dialog appears with a checklist of common hotel items as shown in Figure~\ref{fig:3versions-b} (e.g., bed, lamp, sink). For each image captured, the user identifies the visible items from the list and also has the option to indicate other items. In this case, the collected metadata includes the list of visible objects, but no position information. The WL version results in a 
collection of weakly labeled (i.e., only object names) images. 

\paragraph{Strongly Labeled (SL)}  The user identifies the 
object and location of the object in the photo they take. The camera button is 
part of a swipe-able array of choices corresponding to the same predefined set of hotel objects 
as in WL application. This design was partly inspired by popular camera-based mobile applications (e.g., how a user would choose a face lens or filter in Snapchat). 
When the user selects an item, 
a reticle (i.e., target, bounding box, focus area) is displayed over the view area on the screen outlining an area of interest, as shown in Figure~\ref{fig:3versions-c}. 
 The application provides a default target area for each object, based on the typical 
 size and location of objects. To align objects, the user can choose to (1) resize or 
 move the reticle or 
(2) change locations or the camera angle to bring the object into the target area. The SL version results in images annotated with object names and their locations.

\section{Experiment}
\label{sec:userstudy}

We conducted a study to compare the differences in user contributions between the three interfaces. We ran a between-subjects design with the application version (i.e., labeling level) serving as the independent variable.

\subsection{Study Protocol}
The experiment was carried out in a hotel on campus in one
of two nearly identical hotel rooms. The experimenter provided the following instructions: \textit{``Today, you will serve as a traveler using our mobile application in this hotel. Follow the instructions in the application. When you're done, meet me at [location] and I'll collect your feedback.''} Participants were provided with a smartphone with one of the (randomly-selected) variants of the application pre-loaded. The participants were not provided any limits, requirements, or suggestions on time nor quantity of photographs (or objects) to capture. Immediately after image capture, the investigator collected the smartphone and participants were directed to a laptop in the hotel lobby to complete a brief survey of the experience. The entire experiment session could be completed in less than 5 minutes.

\subsection{Participants}
We recruited participants primarily from a university setting via word-of-mouth by researchers outside of the on-campus hotel. Participants were required to be over 18 years old, have normal 
or corrected-to-normal vision, and be able to read and understand English.
A total of 100 people were recruited to participate in the study (49 male, 51 female). The mean 
age was 21.85 (\textit{SD} = 4.09). 

We asked participants to rate how often they use camera-based smartphone 
applications (e.g., Snapchat, Instagram, etc.) on a scale of 1 = ``Never/Rarely'' 
to 7 = ``Often'' and whether 
they were familiar with the application.  On average, this cohort of mainly 
college-aged participants self-rated as highly familiar with camera-based 
smartphone applications (\textit{M} = 6.25, \textit{SD} = 1.42). The vast majority 
of participants 
(83 out of 100) had not heard of the project. Of those who had, none 
had previously used any version of the mobile application. 
Though the actual application is voluntary, participants in the user study were compensated for their 
time with a \$5 gift card to a campus coffee shop.

\subsection{Data}
We measured user interactions with the application, user satisfaction ratings, and image composition and annotation quality across the three application variants. 

\subsubsection{User Interactions}
For each participant, an event log was recorded, detailing each action performed during the study. Recorded actions included capturing or deleting images, using help screens, and annotating the images. We also recorded how long participants spent using the application. Due to a technical error, the event log from one participant was corrupted, so it is excluded from event-based analysis. 

\subsubsection{User Satisfaction Ratings}
We measured user satisfaction through a post-experiment survey. In addition to gathering demographic information, the questions included rating the interface on a 7-point Likert-type scale on the following criteria: overall quality, instruction quality, and likelihood to recommend to others. 

\subsubsection{Image Composition \& Annotation Quality}
\label{sec:annotators}
All of the images captured during the experiment were evaluated by three annotators. For each image, the annotator marked a bounding box around visible objects from the predefined list. Ground truth annotations were defined as those where at least two annotators agreed on the classification and the bounding boxes significantly overlapped (i.e., Intersection-over-Union (IoU) $>$ .7). The annotators were only provided the captured image and were unaware of application variant used for capture.

\section{Results}
We compared our measures across the three conditions (UL, WL, and SL).
 We tested the normality of our data using the Shapiro-Wilk test. Because our data was non-normal, we used the Kruskal-Wallis test to compare the three conditions on survey and log data and the  multiple comparison test after the Kruskal-Wallis test for post-hoc comparisons on significant results. We used epsilon squared for the effect size~\citep{tomczak2014need}.

\begin{figure}
    \centering
    \includegraphics[width=.75\linewidth]{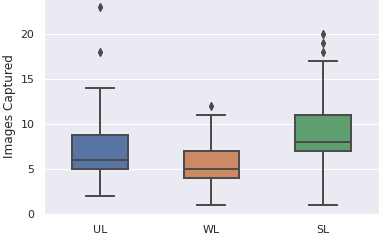}
        \includegraphics[width=.75\linewidth]{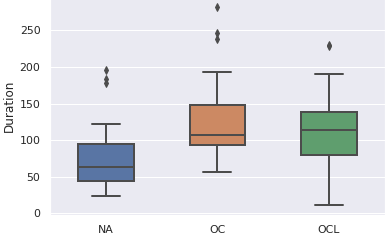}
    \caption{(top) Number of images and (bottom) time in seconds spent capturing (and labeling) images using each application variant.}
    \label{fig:numAndTime}
\end{figure}

\subsection{Number of Photos Taken}
Figure~\ref{fig:numAndTime} (top) shows the number of pictures captured across the three application variants.
We found that participants took the most pictures with the 
SL application variant. 
We found an overall significant difference between the three application versions 
for the number of pictures taken (H(2) = 18.63, $p < 0.001$) with a medium effect size ($\epsilon^2 = 0.19$). 
Because we found a significant difference across the three conditions, we did a post-hoc follow-up test. 
Comparisons of the mean ranks between groups showed that there was a 
significant difference ($p < 0.05$) in the 
number of pictures taken between the UL ($\textrm{Mean} \left(M\right) = 7.44, \textrm{Standard Deviation} \left(SD\right) = 4.5$) and SL conditions ($M = 5.8, SD = 2.67$) (difference = 20.19, critical difference = 16.8). There was also a significant difference ($p < 0.05$) between the 
WL and SL conditions ($M = 9.67, SD = 4.38$) (difference = 29.92, critical difference = 17.06). 

\subsection{Task Times}
Figure~\ref{fig:numAndTime} (bottom) shows the total time the participants spent capturing and/or labeling the image. The duration includes the time spent capturing and labeling images after viewing the instructions. We found a significant difference between task times for the different application versions 
(H(2) = 19.54, $p < 0.001$) with a medium effect size ($\epsilon^2 = 0.2$). Participants spent on average 
75.58 seconds in the UL condition ($SD = 44.42s$), 125.08 seconds in the WL condition ($SD = 55.52s$), 
and 116.07 seconds in the SL condition ($SD = 50.41s$). In the WL condition, we are able to distinguish 
the amount of time participants spent labeling the images because the capturing and labeling
activities are mutually exclusive. Participants in the WL condition spent on average 62.04 seconds labeling ($SD = 23s$), which is almost half of their total task time. Because we found a significant difference across the three conditions, we did a post-hoc follow-up test. Comparisons of the mean ranks 
between groups showed that there were significant differences ($p < 0.05$) in task time 
between the UL and WL conditions (difference = 27.98, critical difference = 16.9) and between the 
UL and SL conditions (difference = 25.25, critical difference = 16.6). There was no significant 
time difference between the WL and SL conditions (difference = 2.74). 


\subsection{Satisfaction Ratings}
The users were asked to rate their overall satisfaction with the application, quality of the instructions, and the likelihood 
of recommending the application to others on a 7-point Likert-type scale. We found no significant difference between 
conditions on participants' overall application rating (H(2) = 5.71, p = 0.057), rating of the instructions (H(2) = 2.76, p = 0.25), 
or recommendation for the application(H(2) = 3.39, p = 0.18). Overall, participants provided positive ratings of their overall 
satisfaction ($M = 5.5, SD = 1.1$), instruction quality ($M = 5.76, SD = 1.35$), and likelihood of recommending the application ($M = 5.78, SD = 1.45$).

\newlength{\exPicLength}
\setlength{\exPicLength}{.12\textwidth}
\begin{figure*}
     \begin{subfigure}{.32\textwidth}
     \centering
    \includegraphics[width=\exPicLength]{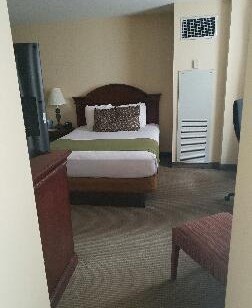}
    \includegraphics[width=\exPicLength]{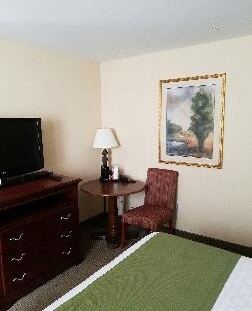}  \\
    \includegraphics[width=\exPicLength]{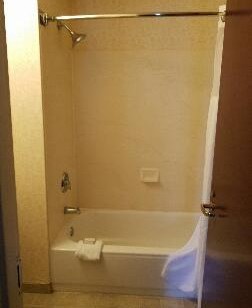}
       \includegraphics[width=\exPicLength]{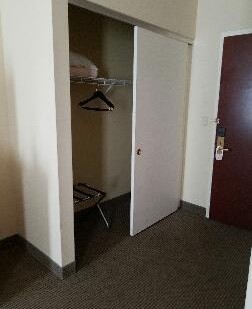}
      \caption{Unlabeled (UL)}
    \end{subfigure}
     \begin{subfigure}{.32\textwidth}
     \centering
    \includegraphics[width=\exPicLength]{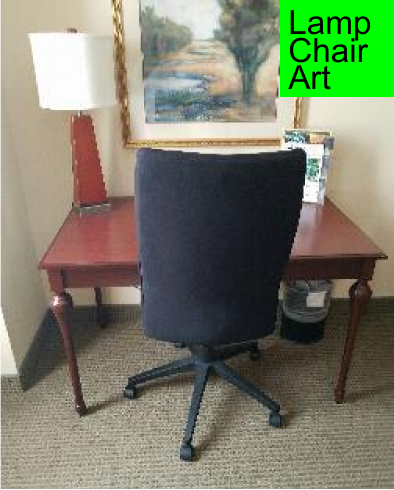}
    \includegraphics[width=\exPicLength]{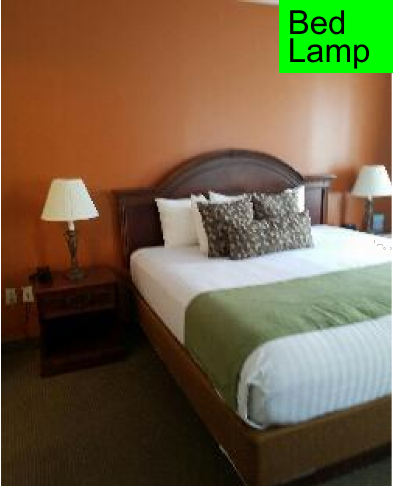} \\ \includegraphics[width=\exPicLength]{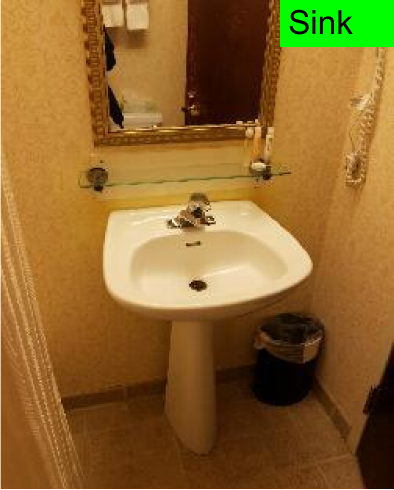}
    \includegraphics[width=\exPicLength]{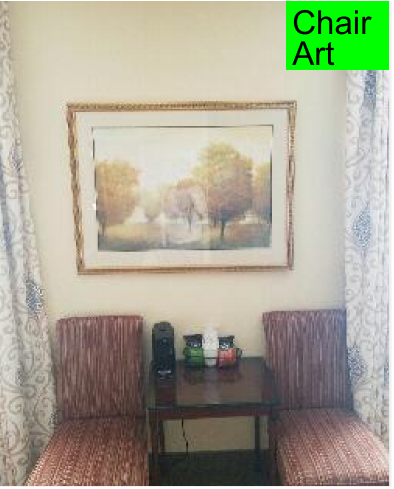}
      \caption{Weakly Labeled (WL)}
    \end{subfigure}
     \begin{subfigure}{.32\textwidth}
     \centering
   \includegraphics[width=\exPicLength]{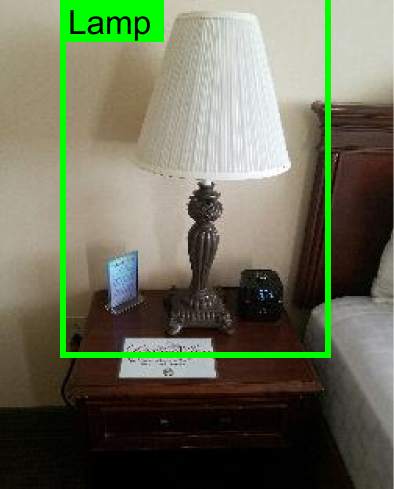}
    \includegraphics[width=\exPicLength]{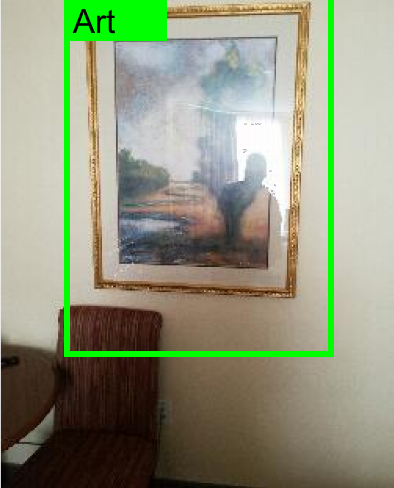} \\ \includegraphics[width=\exPicLength]{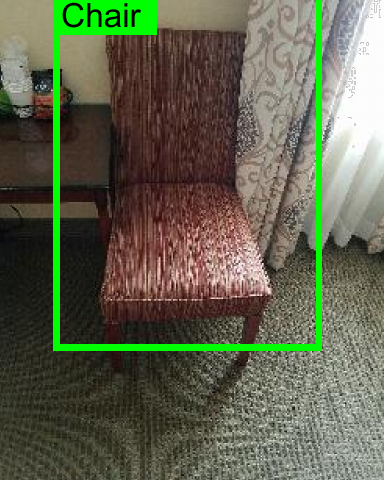}
    \includegraphics[width=\exPicLength]{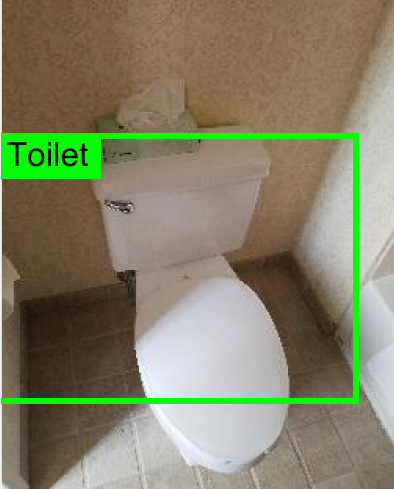}
      \caption{Strongly Labeled (SL)}
    \end{subfigure}

    \caption{Example images captured using three design variants of the camera-based mobile application for capturing images of hotel rooms and identifying objects in the scene.}
    \label{fig:sampleImages}
\end{figure*}

\subsection{Image Composition}
{In addition to understanding whether users are motivated to take pictures, 
crowdsourcing applications rely on their users providing useful and relevant data. 
For this application, the goal is to understand how the different annotation paradigms 
changed the types of pictures taken.} 
Figure~\ref{fig:sampleImages} {shows sample images captured by the study participants 
using each application variant. For the Unlabeled (UL) case, only the image is captured. 
For the Weakly Labeled (WL) case, the inset shows the labels of objects selected by 
the user. For the Strongly Labeled (SL) case, the label and bounding box correspond to the settings selected by the user.
Qualitatively, some visual differences can be observed in the size and positioning of objects
based on whether or not the instructions explicitly referenced objects. To quantify
this phenomenon, we computed both the number of visible objects captured and the 
relative size of those objects in the image frame to serve as proxy measures
for image composition. For all the images, across the three conditions, we used 
the ground-truth annotations provided by the external annotators} (Section~\ref{sec:annotators}) to compute 
the image composition measures.

\begin{figure}
    \centering
        \includegraphics[width=.93\linewidth]{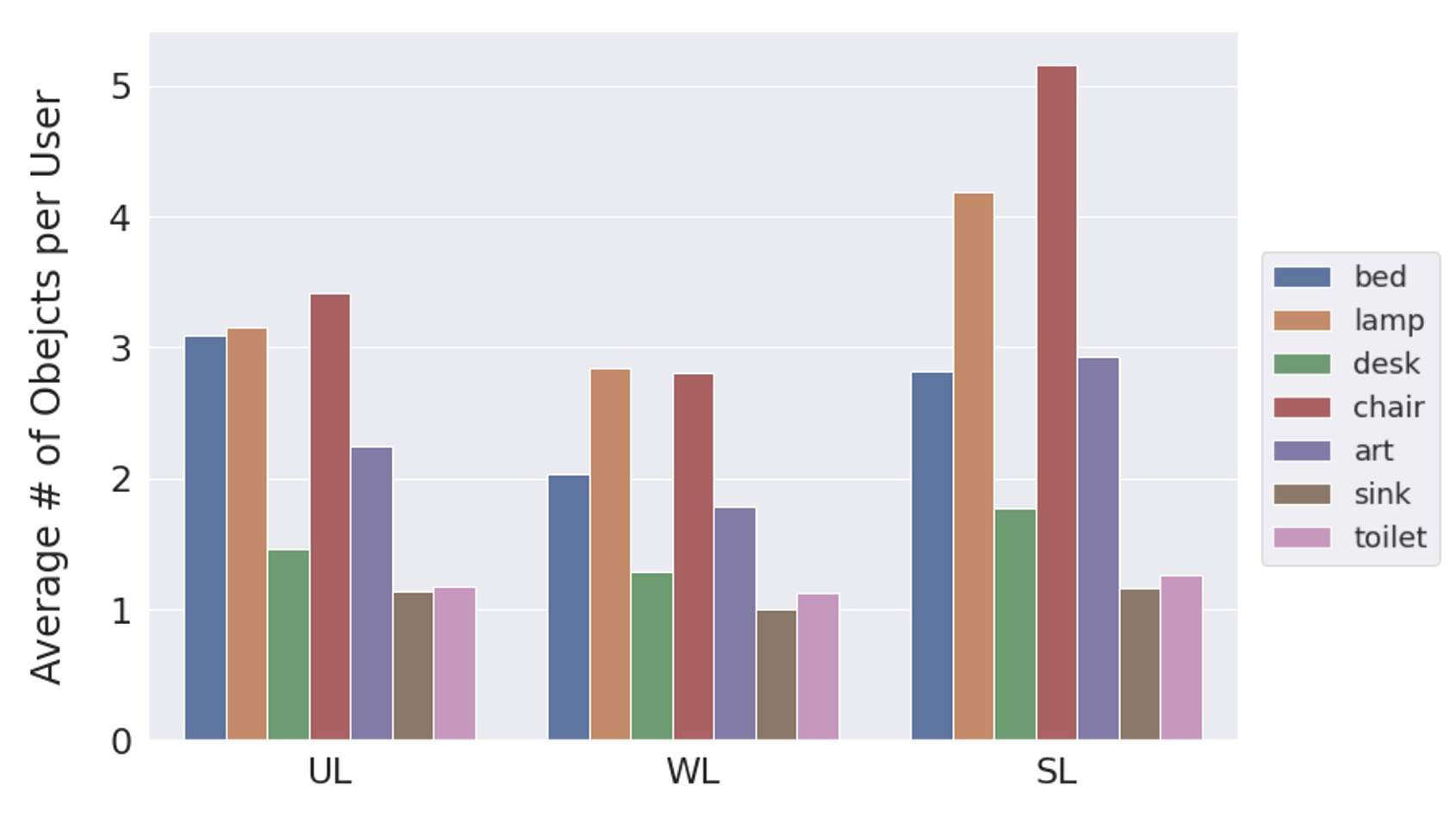}
        \caption{Average number of objects photographed per user.}
    \label{fig:object_counts_sizes}
\end{figure}

The plot in Figure~\ref{fig:object_counts_sizes} shows the average number of hotel objects (from the predefined list) 
photographed per user by application type. In addition to having taken the most images, 
the SL users also captured the most instances of different object types. 
In particular, there is a large increase in the number of lamps and chairs photographed. 

We also computed the average fraction of the image that each object takes up as a function of application type. It is unsurprising that users of the WL and SL application versions capture images with more pixels on the objects 
in general. This supports the finding that the pictures in these conditions were more focused on specific objects, rather 
than pictures in the UL condition that focused more on the overall scene. We found that the fraction of the 
images taken up by objects differed significantly with medium to large effect sizes across conditions for all objects 
except bed, as shown in Table~\ref{table:fractionOfObject2}. This result for beds is reasonable because the relative size of a bed in an image when standing in a small 
hotel room is tightly bounded by the limited amount of space to obtain viewpoints of a large object from different distances.

\begin{table}
\caption{Analysis of the fraction of images occupied by different objects across the three conditions (UL, WL, SL). The
2$^{\text{nd}}$ column shows the results of the Kruskal-Wallis test (H(2), p). The next three columns indicate the pairwise post-hoc comparison differences at $p < 0.05$, and the last column gives the effect size ($\epsilon^2$).}
\label{table:fractionOfObject2}
\centering
\begin{tabular}{|l|c|c|c|c|c|}
\hline
Object & K-W & UL-WL & UL-SL & WL-SL & Eff.\\
\hline
bed & $5.22, p=0.07$ & -- & -- & -- & --   \\
lamp & $45.79, p<0.001$ & \checkmark & \checkmark & \checkmark  &  0.14\\
desk &$27.32, p<0.001$  &  & \checkmark & \checkmark & 0.23 \\
chair &$70.97, p<0.001$ & \checkmark & \checkmark & \checkmark  & 0.2  \\
art &$42.49, p<0.001$  & \checkmark & \checkmark &  & 0.22 \\
sink & $26.65, p<0.001$ & \checkmark & \checkmark & \checkmark  & 0.36   \\
toilet & $37.3, p<0.001$ &  & \checkmark & \checkmark  & 0.43 \\
\hline
\end{tabular}

\end{table}

\section{Discussion}
The goal of this study was to better understand how variations in the design of a camera-centric application can affect the quantity and quality of images and annotations captured by users and user satisfaction with using the application. Towards this end, we evaluated (1) user engagement, as measured by the number of images captured and time spent using the application, (2) the properties of the collected data, and (3) user satisfaction based on the ratings of their experience using the application. In this section, we provide our observations from the results and discuss potential limitations of our experiment. 

\subsection{User Engagement}
The aim of crowdsourcing applications is to collect as much (high-quality) data as 
possible. For this scenario, that translates to users choosing to capture more images. 
The biggest, and most surprising, take-away from this study is that users in SL condition,
which required the most effort and time for image annotation, captured the most images with
the highest variety of objects. Moreover, there was no significant difference 
in the 
user satisfaction ratings or willingness to recommend the application to others, even
though they spent more time, on average, at the task. While we do not take these results to 
imply that the users were equally satisfied across the three conditions, it is noteworthy that 
we did not observe a significant negative correlation between the requested effort and user satisfaction.
These results support the notion that for a crowdsourcing task where the users 
are intrinsically motivated, following the mantra of simplifying the level of 
effort required at the expense of obtaining a higher quantity or quality of data 
may not be warranted.

\subsection{Properties of Collected Data}
The annotation quality results indicate that, for users in this study, high-quality
data can be obtained across annotation paradigms. Downstream algorithms for
computer vision and machine learning only benefit from additional classification
and/or localization annotations in user-provided data. However, it is
important to note that changing the annotation paradigm affected the type
of images that users captured. We observed differences in both the number of objects captured
and the relative sizes of those objects in the image. There were differences among
all three conditions, with UL and WL showing similarity for capturing larger objects (e.g., bed, toilet)
and SL showing differences across most of the object classes compared to both UL and WL.
One possible explanation is that the UL version does not reference objects at all and
the WL version only requests object identification after the image has been captured.
However, for the SL version, object selection and positioning is a part of the image
capture process. It is possible that the indirect prompting inherent to the SL variant
encouraged more photographs of less conspicuous objects (e.g., lamps and chairs) and
zooming in objects that would otherwise be in the background (e.g., art). 

For camera-centric applications, these types of changes to the annotation paradigm may induce unintended 
changes to type of data collected that should be considered by application 
designers. 
For example, converting a camera-centric mobile application from collecting 
unlabeled to strongly labeled data may affect the visual appearance of ``new'' data compared to
previously collected data and require interventions for the downstream
algorithms. 

\subsection{User Feedback}
In addition the numeric ratings in the post-experiment user survey, the participants could also provide free-form comments. Of the 100 participants, 44 left extra comments. Notable themes in the comments include (1) instruction clarity, (2) application intuitiveness, and (3) problem domain interest.

\subsubsection{Instructions}
Although the overall ratings of instructions were high, some users expressed dissatisfaction with the level of detail. In the free response portion of the post-experiment survey, 8 of the 44 users who provided free-form comments mentioned issues with the instructions, with the term `unclear' appearing most frequently in the comments. These comments were split relatively evenly across conditions: 2 for UL, 3 for WL, and 3 for SL. There were also comments that did not call out the instructions specifically, but mentioned not being sure exactly what to take pictures of, such as one participant who commented, \textit{``I wasn't sure if I was supposed to take specific pictures or what exactly would be most helpful.''} Even if the lack of a detail is purposeful to avoid introducing bias, the style and content of the instructions are important design considerations that may confuse or frustrate users and impact continued participation in the crowdsourcing campaign.

\subsubsection{Application Intuitiveness}
While the application design was inspired by popular camera-centric mobile applications, there were implementation choices made to accommodate the annotation tools. Some of these differences were noted by users. Comments include a lack of access to the flash, zoom, and focus controls typically available with camera-based applications. One user additionally desired to ability to re-take their photo, noting \textit{``the picture took blurry a few times, and i couldn't retake a pic.''} In the interest of reducing clutter and to capture click and swipe events related to image annotation, advanced camera controls (e.g., ``pinch'' to zoom) were not enabled, but their inclusion may improve image quality and align with users' expectations of a camera-centric mobile application.

\subsubsection{Problem Domain Interest}
Unlike commercial camera-centric mobile applications, crowdsourcing applications 
can benefit 
from user interest in the problem domain and desire to contribute. Multiple 
participants in this study 
expressed a desire for the application to share more information about the problem 
domain 
and how their contributions help the cause. One user stated, \textit{``I want to know more or be provided 
about the current progress of human tracking through the app. I want to know I am contributing 
to good cause.''} These comments reinforce the notion that, depending on the problem domain,
contributors are often eager to volunteer their time and effort. Providing additional background 
information on the problem domain and, where possible, feedback on the impact of the user's
contribution can encourage continued (and enthusiastic) use of the application.

\subsection{Limitations} 
We identified several threats to validity based on the design of the experiment
and implementation choices.

\subsubsection{Participant Population}
While no participants had ever used any version of the application, 
some participants were familiar with the project 
through contact with some of the researchers (students, professors). 
Participants were also compensated with a \$5 gift card. 
Either of these factors could lead to inflated ratings. While participants gave a 
similar overall rating whether they were familiar ($M = 5.5, SD = 0.97$) or not familiar with the 
project ($M = 5.51, SD = 1.09$), they did score their willingness to recommend the application slightly higher if they had familiarity ($M = 5.94, SD = 1.7$) than if they did not ($M = 5.75, SD = 1.4$). 
Participants familiar with the application were relatively well spread across conditions: 
5 of them used the UL version, 7 used the WL version and 4 of them used the SL condition.

Additionally, the participants consisted primarily of undergraduate students. 
This population may have more time and 
motivation to participate in this type of crowdsourcing, and the application, which 
is intended to help combat human trafficking, may inspire a higher level of altruism due to the 
subject matter. However, we expect that all of these factors would have affected the population 
overall, rather than one condition or application type.

\subsubsection{Variability of Labeling Implementations}
The UL variant simply requires capturing images and adds little functionality beyond the built-in camera application common to
all smartphones. There are few design decisions to be made. However, the WL and SL variants fall into broader categories where choices such as the order of labeling and image capture and/or the number of objects to label per image can affect the design and implementation.
For the WL variant, we are reasonably confident that it aligns
with other applications in the weakly labeled paradigm and the different interaction modes (e.g., \emph{label-then-capture} vs. \emph{capture-then-label}) only constitute minor differences. However, as
previously mentioned, the strongly labeled paradigm is much broader. Applications
fitting this paradigm have employed opaque overlays, reticles or target areas (as with SL),
and other interaction widgets. Aligning an object in a scene with a marker in the viewfinder
can be accomplished by manipulating the marker or capturing the image from a different angle. 
Additionally, some approaches rely on bounding boxes while others employ tools for pixel-segmentation
of objects. While we aimed to provide some amount of flexibility in our SL implementation (e.g., resizable reticle),
the results may not generalize to the wide variety of approaches for strongly labeling images.

\subsubsection{Environment Constraints}
Two additional limitations of this study were the fact that images were only captured in
one of two different (but nearly identical) rooms from the same hotel and the availability of objects did
not include all of the objects that might be expected. For example, 
neither room contained a couch. Nonetheless, we expect that the general measurements of 
user effort would not be significantly 
impacted by either of these issues.

\subsection{Labeling Alternatives}
Rather than redesigning an existing (most likely unlabeled) camera-centric mobile application, one might consider alternatives for obtaining image labels. One option
would be to crowdsource the labeling task after the images have been captured 
on a platform like Amazon Mechanical Turk. This option introduces additional 
costs and 
leaves the task to a different set of users who may not be as motivated as the 
cohort that captured the images.  Another approach to labeling involves the use of automated algorithms. While these methods are close to human-level performance, even state-of-the-art approaches still misidentify or oversegment objects. It is worth noting that our task involves objects (e.g., bed, chair) commonly included in the generic data sets used to train these methods. Even in this case, one of the automated methods did not include relatively common items (e.g., lamp, desk). This issue is only exacerbated for the specialized tasks for which camera-based crowdsourcing has been employed (e.g., litter, potholes, fine-grained animal or plant species recognition), limiting the utility of automated labeling approaches in these domains.
One last consideration for both of these alternatives is that our results show that 
by mentioning the labeling task during image capture, users take more images and more images focused 
on specific objects, which may or may not be desirable, depending on the task.

\subsection{Real-World Deployment}
\label{sec:production}

\begin{figure}
    \centering
    \includegraphics[width=.9\linewidth]{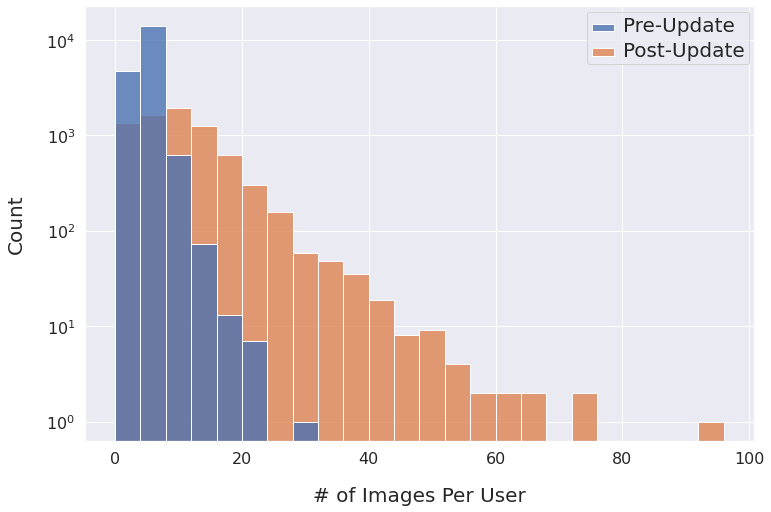}
    \caption{Distribution of the number of images per user before (blue) and after (orange) introducing the strong labeling interface. 
    }
    \label{fig:production_app_stats}
\end{figure}

Based on the results of the user study, the application was updated to incorporate the strongly labeled paradigm. Users could opt to submit images in the original manner (unlabeled, images only) or
a new mode similar to the SL paradigm with the object carousel and object reticle. Prior to the update, users provided 3.68 ($SD~=~1.49$) images on average. Since the update, users have submitted an average of 10.06 ($SD~=~7.49$) images. These distributions are shown in Figure~\ref{fig:production_app_stats}. As the update introduced many changes, including the UI, instructions, and ease of uploading, it would not be fair to attribute this increase entirely to introducing the object-centric SL paradigm option. It is worth noting that users have the option to choose between providing unlabeled or labeled images, and 88.4\% provided more strongly labeled images than unlabeled images. 

\section{Object-Centric Image Recognition}
The data collected by the crowdworkers can be used to train automated methods for fine-grained  recognition. Hotels-50K~\citep{hotels50k} is a benchmark dataset of more than a million images 
from 50,000 hotels around the world. This hotel recognition task in this benchmark closely resembles the investigative task in human trafficking investigations, and the benchmark includes all of the challenges common to general fine-grained categorization and
others unique to hotel rooms. 
There can be a high within-class variation; images from rooms from the same hotel can appear to be quite different. Also, there can be low between-class variation, particularly for images from hotels belonging to the same chain. 

\newlength{\myheight}
\setlength{\myheight}{2.47cm}

\begin{figure}

\begin{subfigure}[b]{.46\textwidth}
\centering
\includegraphics[height=\myheight]{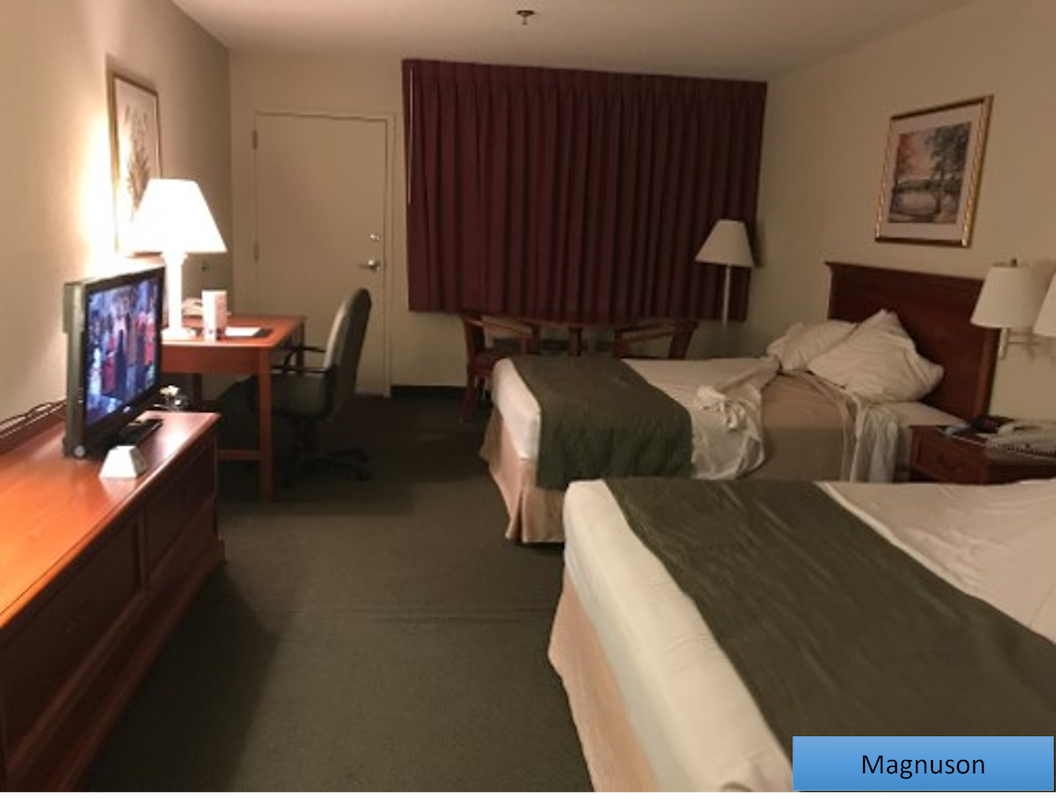}
\includegraphics[height=\myheight]{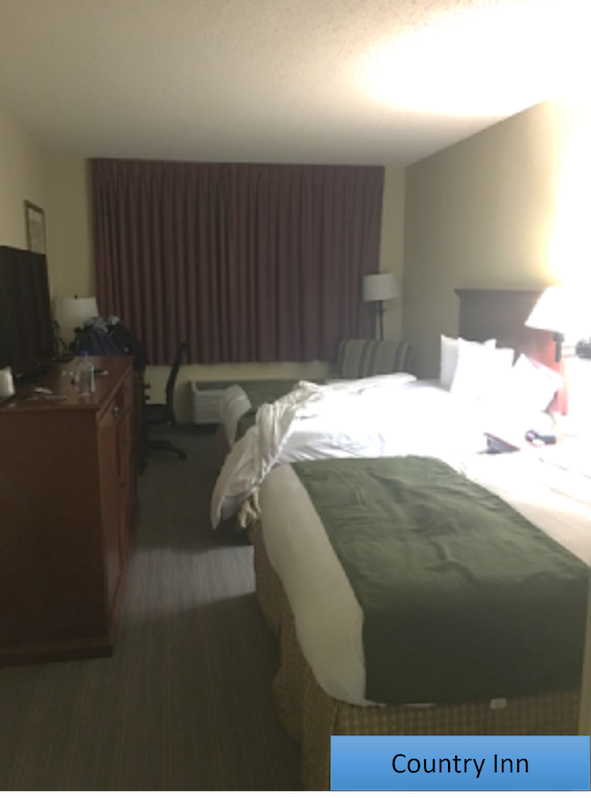}
 \caption{similar scene, different objects}
  \label{fig:failure-cases1}
\end{subfigure}
\begin{subfigure}[b]{.46\textwidth}
\centering
\includegraphics[height=\myheight]{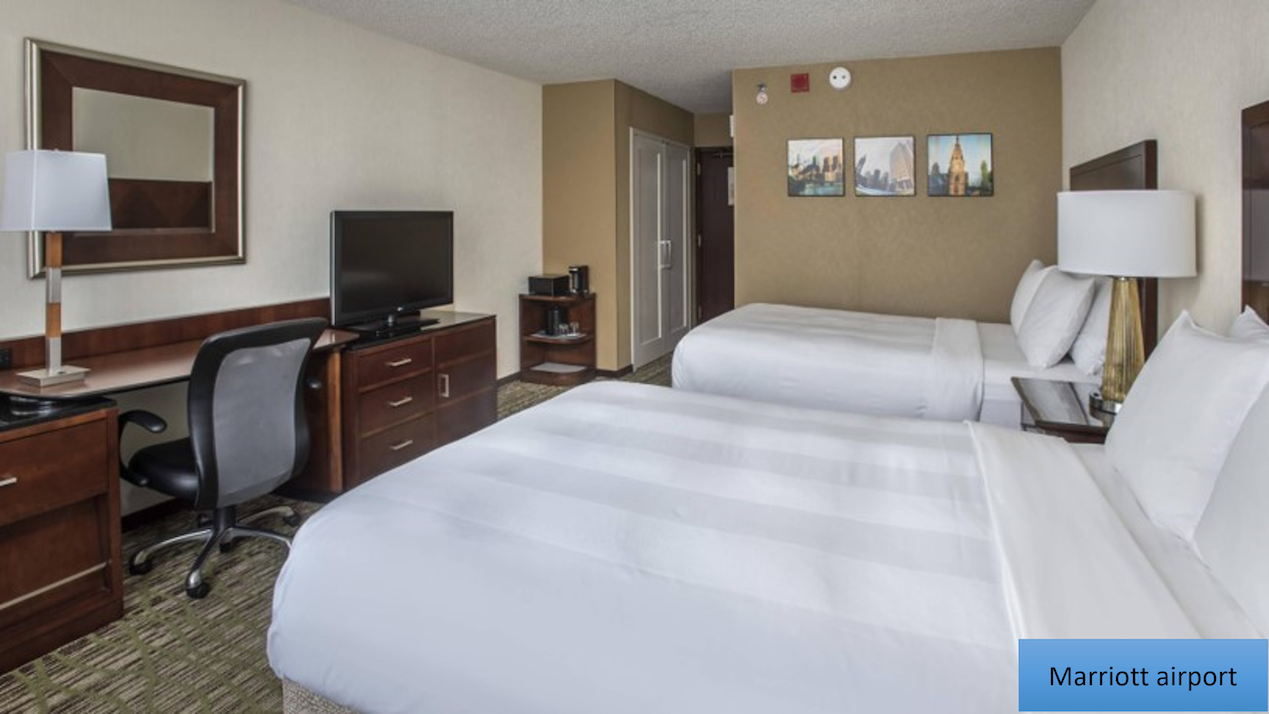}
 \includegraphics[height=\myheight]{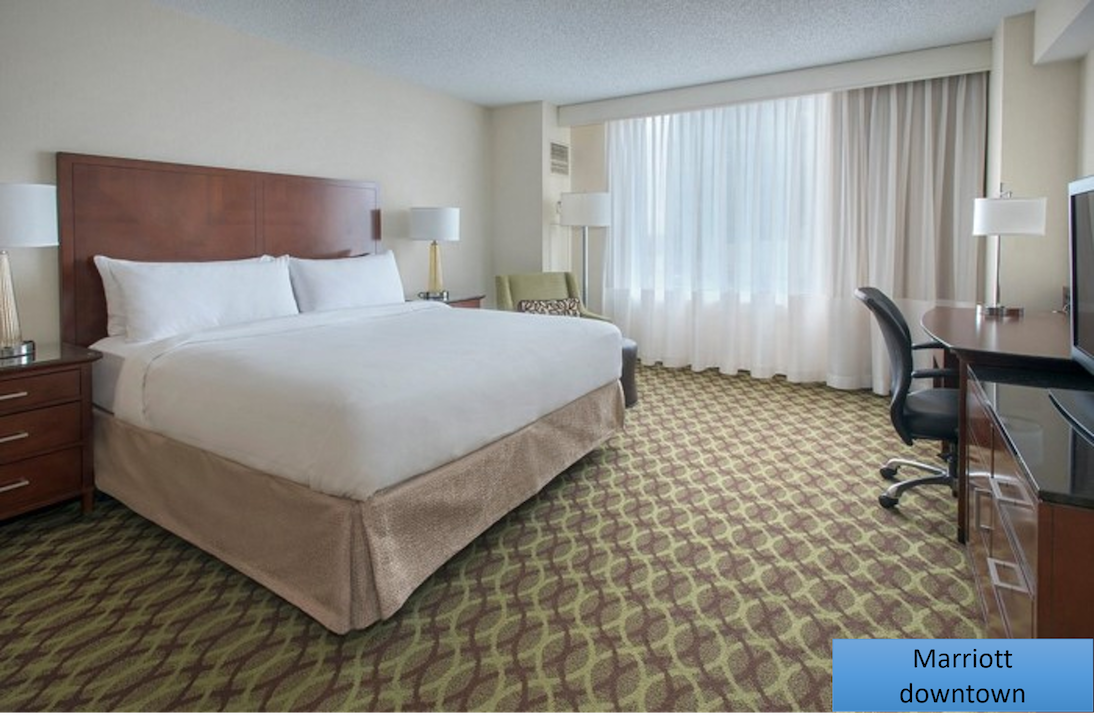}
 \caption{different scene, similar objects}
 \label{fig:failure-cases2}
\end{subfigure}
  \caption{(a) Images representing different hotels that are visually similar, but close inspection of the objects (lamps, chairs, dressers) show clear differences. (b) Images from the same hotel chain that are visually dissimilar but contain similar objects (lamps, chairs).}
    \label{fig:imagevsobject}

\end{figure}

The objects visible in the hotel room images can help discriminate between rooms. Previous approaches focused on image-level matching, treating the entire image as input to the model. Figure~\ref{fig:imagevsobject} highlights
the drawbacks of this approach; in (a), the two images, although visually quite similar, were captured from different hotels. Closer inspection of some of the objects (e.g., lamps, chairs, dressers) shows clear differences. In (b), the images are from the same hotel chain and contain the same objects (lamps, chairs), but the overall scenes are visually quite dissimilar. 

As an alternative to image-centric approaches, we take an object-centric approach by representing the image as 
a set of objects (e.g., lamps, beds, curtains) and applying a simple voting scheme for matching to a database of object crops. Using a ViT/S with 16x16 patch size~\cite{DosovitskiyB0WZ21}, we trained an image-based classifier, achieving a baseline accuracy of .551.
Using the same model, we extracted the image patches corresponding to the specific objects
in the scene; a patch token is associated with the object that occupies the majority of pixels within that patch. These patches were mean-pooled to serve as the object feature, and we used a majority-rule scheme to aggregate the predictions of all the objects in the image. This simple object-based strategy achieved an accuracy of .583, outperforming the standard image-based strategy by over 11\%. This experiment highlights the benefit of the localized, labeled data provided by the crowdsourced users.

\section{Conclusion}
This paper contributes one of the first investigations into the trade-offs between labeling effort, 
quality and quantity of data collected, and user satisfaction for a camera-centric 
crowdsourcing application. The results suggested that motivated users may be willing
to do more than universally accepted design guidelines (i.e., ``keep it simple'')
may suggest. Even in this limited study of three variants of an application for a particular task, 
we observed a complex, interconnected relationship between the design choices, user expectations, 
and user behavior. Finally, we demonstrated that data from a real world camera-centric crowdsourcing application can be used to improve image retrieval performance. 

\section{Acknowledgments}
The work described in this paper was supported by National Institute of Justice Award \#2018-75-CX-0038 and National Science Foundation Award \#1757533. We greatly appreciate the work of Gabe Aguilar and Madilyn Simons for developing the version of the mobile application used in the study and Nick Allan for drawing his illustrations for both the instructions and figures in this paper. We would also like to thank Jake Lawrence, David Zheng, and Kat Osadchuk for helping recruit participants during the user study.

\bibliographystyle{ACM-Reference-Format}
\bibliography{main}

\end{document}